\documentclass[a4paper,prl,noshowpacs,noshowkeys,twocolumn]{revtex4}
\usepackage{epsfig,amsfonts,amsthm}
\newcommand{\be}{\begin{equation}}
\newcommand{\ee}{\end{equation}}
\newcommand{\bea}{\begin{eqnarray}}
\newcommand{\eea}{\end{eqnarray}}
\newcommand{\dst}{\displaystyle}
\newcommand{\fr}[2]{\frac{{\dst #1}}{{\dst #2}}}

\renewcommand{\Re}{\mbox{Re }}
\renewcommand{\Im}{\mbox{Im }}
\newcommand{\stolb}[3]{ \left( \begin{array}{c}#1 \\ #2 \\ #3\end{array}\right) }

\newcommand{\lr}[1]{ \langle #1 \rangle}

\newtheorem{theorem}{Theorem}
\newtheorem{proposition}[theorem]{Proposition}

\def\lsim{\mathrel{\rlap{\lower4pt\hbox{\hskip1pt$\sim$}}
    \raise1pt\hbox{$<$}}}         
\def\gsim{\mathrel{\rlap{\lower4pt\hbox{\hskip1pt$\sim$}}
    \raise1pt\hbox{$>$}}}         

\begin{document}

\title{Geometric methods for the most general Ginzburg-Landau model with two order parameters}

\author{I.~P.~Ivanov}
\email{Igor.Ivanov@ulg.ac.be}
\affiliation{
  Interactions Fondamentales en Physique et en Astrophysique, Universit\'{e} de Li\`{e}ge, \\
  All\'{e}e du 6 Ao\^{u}t 17, b\^{a}timent B5a, B-4000 Li\`{e}ge, Belgium
}
\affiliation{
Sobolev Institute of Mathematics, Koptyug avenue 4, 630090, Novosibirsk, Russia}

\begin{abstract}
The Landau potential in the general Ginzburg-Landau theory with two order parameters and all possible
quadratic and quartic terms cannot be minimized with the straightforward algebra.
Here, a geometric approach is presented that circumvents this computational difficulty
and allows one to get insight into many properties of the model in the mean-field approximation.
\end{abstract}

\maketitle

{\em Introduction.}
The Ginzburg-Landau (GL) theory offers a remarkably economic description
of phase transitions associated with breaking of some symmetry \cite{GL}.
This breaking is described
with an order parameter $\psi$: the high symmetry phase corresponds to
$\psi=0$, while the low symmetry phase is described by
$\psi \not =0$.
In order to find when a given system is in high or low symmetry
phase, one constructs a Landau potential
that depends on the order parameter, and then find its minimum.
Its classic form is
\be
V(\psi) = - a |\psi|^2 + {b \over 2} |\psi|^4 + o(|\psi|^4)\,.\label{landau}
\ee
Higher order terms $o(|\psi|^4)$ are usually assumed to be negligible.
The values of the coefficients $a$ and $b$ and their dependence on temperature, pressure, etc.
can be either calculated from a microscopic theory, if it is available,
or considered as free parameters in a phenomenological approach.
The phase transition associated with the symmetry breaking takes place
when an initially negative $a$ becomes positive, and the minimum of the potential (\ref{landau})
shifts from zero to $|\psi| = \sqrt{a/b}$.

Many systems are known, in which two competing order parameters (OP) coexist.
Among them are general $O(m)\oplus O(n)$-symmetric models, \cite{EXOmOn},
models with two interacting $N$-vector OPs with $O(N)$ symmetry, \cite{EXtwoOn};
spin-density-waves in cuprates, \cite{EXspindensity};
$SO(5)$ theory of antiferromagnetism and superconductivity, \cite{EXantiferrosuper};
multicomponent, \cite{EXmulticomponentsuper}, spin-triplet $p$-wave, \cite{EXtripletsuper},
and two-gap, \cite{EXtwoband,EXtwogapGurevich}, superconductivity,
with its application to magnetism in neutron stars, \cite{EXneutronstars};
two-band superfluidity, \cite{EXtwobandfluid};
various mechanisms of spontaneous breaking of the electroweak symmetry
beyond the Standard Model such as the two-Higgs-doublet model (2HDM), \cite{EX2HDM}.

To describe such a situation within the GL model,
one constructs a Landau potential similar to (\ref{landau}),
which depends on two order parameters, $\psi_1$ and $\psi_2$.
Coefficients of this potential, $a_i$ and $b_i$, can be
considered independent, although in each particular
application they might obey specific relations.
One thus arrives at the most general two-order-parameter (2OP) GL model
with quadratic and quartic terms.

A natural question arises: what is the ground state of the most general
2OP GL model?
A rather surprising fact is that this question cannot be answered
by a straightforward calculation. Differentiating the Landau potential
in respect to $\psi_i$ leads to a system of coupled algebraic
equations that cannot be solved explicitly.

In this Letter we show that despite this computational difficulty, one
can still learn much about the most general 2OP GL model.
Namely, one can study the number and the properties of the
minima of the Landau potential, classify possible symmetries
and study when and how they are broken.
In short, one can describe the phase diagram of the model,
at least in the mean-field approximation,
without explicitly minimizing the potential.

There exists, in fact, an extensive literature dating back to 1970's
on minimization of group-invariant potentials with several OPs
with the aid of stratification of the orbit space,
see e.g. \cite{GL,sartori}.
Here we show that in the case of two order parameters
realizing the same group representation
the analysis can be extended much farther than in the general case,
with important physical consequences.
For a particular application of this formalism to the 2HDM, see
\cite{mink,minknew}.

{\em The formalism.}
Let us focus on the simplest case when two OPs $\psi_1$ and $\psi_2$
are just complex numbers.
The most general quadratic plus quartic Landau potential is
\bea
V &=& -a_{1}|\psi_1|^2 - a_{2}|\psi_2|^2
- a_{3} (\psi_1^*\psi_2) - a_{3}^{*} (\psi_2^*\psi_1)\nonumber\\
&+&
\fr{b_1}{2}|\psi_1|^4 + \fr{b_2}{2}|\psi_2|^4
+ b_3 |\psi_1|^2|\psi_2|^2 \label{potential}\\
&+&\left[\fr{b_4}{2}(\psi_1^*\psi_2)+
b_5 |\psi_1|^2 + b_6 |\psi_2|^2\right](\psi_1^*\psi_2) +{\rm
c.c.}\nonumber
\eea
It contains 13 free parameters: real
$a_{1},\, a_{2},\, b_1,\, b_2,\, b_3$ and
complex $a_{3},\, b_4,\, b_5,\, b_6$.
For the illustration of the main idea, we place
no restriction on $|\psi_i|$ from above.
Note that potential (\ref{potential}) contains quartic terms that mix
$\psi_1$ and $\psi_2$. Such terms are usually absent
in particular applications of the 2OP GL model
(for a rare exception, see \cite{EXtwogapGurevich}),
but in the approach presented here it is essential that
all possible terms are included from the very beginning.

Once potential (\ref{potential}) is written,
the physical nature of OPs becomes irrelevant.
One can consider them as components
of a single complex 2-vector $\Phi = (\psi_1, \psi_2)^T$.
The key observation is that the most general potential
(\ref{potential}) keeps its generic form
under any regular linear transformation between $\psi_1$ and $\psi_2$:
$\Phi \to \Phi' = T\cdot \Phi$, $T \in GL(2,C)$.
It can be also accompanied with a suitable
transformation of the coefficients $a_i,\, b_i$, so that one arrives at
exactly the same potential as before.
Thus, the problem has some {\em reparametrization freedom}
with the reparametrization group $GL(2,C)$. Among 13 free parameters,
only 6 play crucial role in shaping the phase diagram of the model,
while the other 7 just reflect the way we look at it.

Let us now introduce a four-vector $r^\mu = (r_0,\,r_i) = (\Phi^\dagger \sigma^\mu \Phi)$ with components
\bea
r_0 &=& (\Phi^\dagger \Phi) = |\psi_1|^2 + |\psi_2|^2\,,\nonumber\\
r_i &=& (\Phi^\dagger \sigma_i \Phi) =
\stolb{2\Re (\psi^*_1 \psi_2) }{2\Im (\psi^*_1 \psi_2) }{|\psi_1|^2 - |\psi_2|^2}\,.
\label{ri}
\eea
Here, index $\mu = 0,1,2,3$ refers to components in the internal space
and has no relation with the space-time.
Multiplying $\psi_i$ by a common phase factor does not change $r^\mu$,
so each $r^\mu$ parametrizes a $U(1)$-orbit in the $\psi_i$-space.
Since the Landau potential is also $U(1)$-invariant,
it can be defined in this $1+3$-dimensional orbit space.

The $SL(2,C) \subset GL(2,C)$ group of transformations of $\Phi$ induces
the proper Lorentz group $SO(1,3)$ of transformations of $r^\mu$.
This group includes 3D rotations of the vector $r_i$ as well as ``boosts''
that mix $r_0$ and $r_i$, so
the orbit space gets naturally equipped with the Minkowski space structure.
Since $r_0>0$ and $r^\mu r_\mu \equiv r_0^2 - \sum r_i^2 = 0$,
the orbit space is given by the ``forward lightcone'' $LC^+$ in the Minkowski space.

All this allows us to rewrite (\ref{potential}) in a very compact form:
\be
V = - A_\mu r^\mu + {1\over 2} B_{\mu\nu} r^\mu r^\nu\,,\label{potential2}
\ee
with
\bea
&&\hspace{-8mm}A^\mu = {1\over 2}\left(a_{1}+a_{2},\, -2\Re a_{3},\,
2\Im a_{3},\, -a_{1}+a_{2}\right)\,,\nonumber\\[2mm]
&&\hspace{-8mm}
B^{\mu\nu} = {1\over 2}\left(\begin{array}{cccc}
{b_{12}^+ \over 2} + b_3 & -\Re b_{56}^+
    & \Im b_{56}^+ & - {b_{12}^- \over 2} \\[1mm]
-\Re b_{56}^+ & \Re b_4 & -\Im b_4 & \Re b_{56}^- \\[1mm]
\Im b_{56}^+ & -\Im b_4 & -\Re b_4 & -\Im b_{56}^- \\[1mm]
 -{b_{12}^- \over 2} & \Re b_{56}^- & -\Im b_{56}^-
 & {b_{12}^+ \over 2} - b_3
\end{array}\right).\label{KAB}
\eea
Here, $b_{12}^\pm \equiv b_1\pm b_2$, $b_{56}^\pm \equiv b_5\pm b_6$.

One usually requires that the quartic term of the potential increases
in all directions in the OP-space.
In the orbit space, this was proved in \cite{mink}
to be equivalent to the statement that $B_{\mu\nu}$ is diagonalizable
by an $SO(1,3)$ transformation and after diagonalization it takes form
$B_{\mu\nu} = \mathrm{diag}(B_0,\, -B_1,\, -B_2,\, -B_3)$
with
\be
B_0 > B_1, B_2, B_3\,.\label{Bi}
\ee
Since $r^\mu r_\mu = 0$, the matrices $B^{\mu\nu}$ and $\tilde{B}^{\mu\nu} = B^{\mu\nu} - C g^{\mu\nu}$
are equivalent. This degree of freedom in the definition of $B^{\mu\nu}$
shifts all the eigenvalues by a common constant.

Finding eigenvalues $B^{\mu\nu}$ explicitly in terms of $a_i$, $b_i$
requires solution of a fourth-order characteristic equation,
which constitutes one of the computational difficulties
of the straightforward algebra.
We reiterate that in our analysis we never use these explicit expressions.
The analysis relies only on the fact that the eigenvalues
are real and satisfy (\ref{Bi}).

{\em Minima of the Landau potential.}
Let us first find how many extrema the potential (\ref{potential2})
can have in the orbit space.
Since extrema lie on the surface of $LC^+$, we use the Lagrange multiplier method
to arrive at the following simultaneous equations:
\be
B_{\mu\nu} \lr{r^\nu} - \lambda \lr{r_\mu} = A_\mu\,,\quad
\lr{r^\mu}\lr{r_\mu} = 0\,.\label{lagrange}
\ee
Here, $\lr{r^\mu}$ labels the position of an extremum.
This system cannot be solved explicitly in the most general case,
however one can establish {\em how many} extrema a given potential has.

\begin{figure}[!htb]
   \centering
\includegraphics[width=4cm]{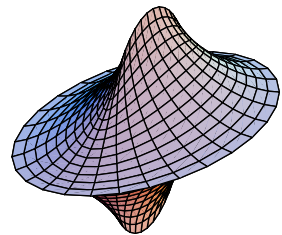}
\caption{The envelope of ellipsoids for $r_0^{(1)}<r_0<r_0^{(2)}$ in the $A_i$-space.}
   \label{fig-caustic-region}
\end{figure}

To find it, we rewrite $\lr{r_i} = \lr{r_0} n_i$, where $n_i$ is a unit 3D vector;
then (\ref{lagrange}) becomes
\be
\left[A_0 - (B_0-B_i)\lr{r_0}\right]n_i = A_i\,.\label{lagrange2}
\ee
Assume for simplicity that $A_0>0$ and $B_1<B_2<B_3$.
The l.h.s. of (\ref{lagrange2}) at fixed $\lr{r_0}$ and all unit vectors $n_i$
parametrizes an ellipsoid with semiaxes $A_0 - (B_0-B_i)\lr{r_0}$.
As $\lr{r_0}$ increases from zero to infinity,
this ellipsoid first shrinks, then grows, collapsing at
$r_0^{(i)} \equiv A_0/(B_0-B_i)$ to planar ellipses.
One can see that during these transformations for $\lr{r_0}=[0,\infty)$
it sweeps at least once the entire Minkowski space
and at least twice the interior of the sphere of radius $A_0$.
In addition, there are two cusped regions, such as shown in Fig.~\ref{fig-caustic-region},
whose interior is swept twice more.
So, by checking whether $A_i$ lies inside these regions, one can get the number of solutions
of (\ref{lagrange2}) without finding them explicitly.

In the $1+3$-dimensional space of $A^\mu$, these 3D regions
serve as bases of corresponding conical regions with different numbers of
extrema of the potential.
Namely, at least one non-trivial solution of (\ref{lagrange}) exists,
if $A^\mu$ lies outside the past lightcone $LC^-$
(otherwise the global minimum of the potential is
at the origin, $\langle\psi_i\rangle = 0$).
If $A^\mu$ lies inside the future lightcone $LC^+$,
then there are at least two non-trivial solutions.
If in addition $A^\mu$ lies inside one or both caustic cones,
then two additional extrema per cone exist.
In total, the potential can have up to six non-trivial extrema in the orbit space.
This fact was also found independently in \cite{nachtmann}.

The above construction does not distinguish a minimum from a saddle point
(with condition (\ref{Bi}), there are no non-trivial maxima in the orbit space).
A straightforward method for finding a minimum, which consists in checking
that the hessian eigenvalues are all non-negative,
is again of little use here.
Instead, one can still use geometry to study the properties of the minima.

As described above, physically realizable points of the orbit space
lie on $LC^+$. Nevertheless, let us consider expression (\ref{potential2})
in the entire Minkowski space. Let us define
an equipotential surface ${\cal M}^V$ as a set of all vectors $r^\mu$
with the same value of $V$.
These equipotential surfaces do not intersect,
are nested into each other, and have very simple
geometry: they are second-order 3-dimensional manifolds (3-quadrics).
Since $LC^+$ is also a specific 3-quadric, finding points in the orbit space
with the same value of $V$ amounts to finding intersections of these two 3-quadrics.
In particular, to find a local minimum of the potential in the orbit space,
one has to find an equipotential surface that touches
$LC^+$ in an isolated point (we say that two surfaces ``touch'' if they have parallel normals
at the intersection points).
The global minimum corresponds to the unique equipotential surface that only touches but never intersects
$LC^+$. The fact that the search for the global minimum is reformulated in terms of
contact of two 3-quadrics leads to several Propositions listed below.

Let us now find how many among the extrema are minima.
Let us fix $B^{\mu\nu}$ and move $A^\mu$ continuously in the parameter space.
We first note that the signature of the hessian can change only when
the total number of extrema changes.
A saddle point cannot simply become a minimum;
it can only split into several extrema, one of them being a minimum,
or it can merge with other extrema to produce a minimum.
Therefore, the conical 3-surfaces described above ($LC^-$, $LC^+$, caustic cones)
separate regions in the $A^\mu$-space with a definite number of minima.
One can then check a representative point $A^\mu = (A_0,\,0,\,0,\,0)$
(in the basis where $B^{\mu\nu}$ is diagonal)
of the innermost region in the $A^\mu$ space
and find that there are two distinct minima in this case.
This proves the following Proposition:
\begin{proposition}\label{prop-two-minima}
The most general quadratic plus quartic potential with two order parameters
can have at most two distinct local minima in the orbit space.
\end{proposition}

{\em Symmetries and their violation.}
The potential can have an additional explicit symmetry,
i.e. it can remain invariant under some transformations
of $\psi_i$ (or coefficients) alone.
If the position of the global minimum is also invariant under the
same group of transformations, we say that the symmetry is preserved;
otherwise, we say that the explicit symmetry is spontaneously violated.
In most applications, the Landau potential does
possess some explicit symmetry, so
whether it is preserved or violated can have profound
physical consequences.

An explicit symmetry corresponds to such a map of the Minkowski space
that leaves invariant, separately, $B_{\mu\nu} r^\mu r^\nu$ and $A_\mu r^\mu$.
In a general situation, it might be far from evident that the potential
has any explicit symmetry. The following criterion
helps recognize the presence of a hidden explicit symmetry and tells
what symmetry it is:
\begin{proposition}\label{prop-classification-symmetries}
Suppose that the potential is explicitly invariant under some
transformations of $r^\mu$.
Let $G$ be the maximal group of such transformations.
Then:\\
(a) $G$ is non-trivial if and only if there exists an eigenvector
of $B_{\mu\nu}$ orthogonal to $A_\mu$;\\
(b) $G$ is one of the following groups: $Z_2$, $(Z_2)^2$,
$(Z_2)^3$, $O(2)$, $O(2)\times Z_2$, $O(3)$.
\end{proposition}
The proof will be given in \cite{GLlong} (see also \cite{group}
for a similar statement in 2HDM).
In the case of a discrete symmetry
the following Propositions can be easily proved:
\begin{proposition}\label{prop-maximal-violation}
The maximal violation of any discrete explicit symmetry consists in removing
only one $Z_2$ factor: $(Z_2)^{k} \to (Z_2)^{k-1}$.
\end{proposition}
\begin{proposition}\label{prop-noncoexistence}
For any explicit discrete symmetry, minima that preserve and
spontaneously violate this symmetry cannot coexist.
\end{proposition}
Both Propositions follow from Proposition~\ref{prop-two-minima}
and the fact that the set of all minima preserves
the explicit symmetry group $G$.

If a discrete symmetry is spontaneously violated,
then there are two generate minima in the orbit space.
One can also prove the converse, i.e. the two degenerate minima
can arise only via spontaneous violation of a discrete symmetry of the
potential, \cite{minknew}.
Thus, the criterion for the spontaneous violation of a discrete
symmetry is that $A_\mu$ lies inside a caustic cone associated
with the largest eigenvalue.

For a concrete example, suppose that all eigenvalues of $B_{\mu\nu}$ are distinct
and that $A_3 = 0$, while other components $A_i \not = 0$.
This potential has an explicit $Z_2$ symmetry generated
by reflections of the third coordinate.
The global minimum spontaneously violates this symmetry (i.e. $\lr{r_3} \not = 0$),
if $B_3 > B_1,\, B_2$ and
\be
{A_1^2 \over (B_3-B_1)^2} + {A_2^2 \over (B_3-B_2)^2} < {A_0^2 \over (B_0-B_3)^2}\,.
\label{violation}
\ee
The proof is based on the ``shrinking ellipsoid'' construction described above and
will be given in detail in \cite{GLlong}.

{\em Local order parameters.}
In this Letter we have illustrated the idea using the global OPs $\psi_i$.
The approach can be easily extended to models, where OPs are defined
locally, $\psi_i(x)$. In this case, one considers the free-energy functional
$F[\psi_i] = \int d^3 x [K(\psi_i) + V(\psi_i)]$ with kinetic term $K$ and potential $V$.
In the general model the kinetic term must include all quadratic combinations of the
gradient terms:
\be
K = \kappa_{1} |\vec{D} \psi_1|^2 + \kappa_{2} |\vec{D} \psi_2|^2
+ \kappa_{3} (\vec{D}\psi_1)^*(\vec{D} \psi_2) + c.c.
\,,\label{gradient}
\ee
where $\vec{D}$ is either $\vec{\nabla}$ or the covariant derivative.
It can be rewritten in the reparametrization invariant form
$K = K_\mu \rho^\mu$ with $\rho^\mu \equiv (\vec{D} \Phi)^\dagger \sigma^\mu (\vec{D} \Phi)$
and $K_\mu$ defined in terms of $\kappa_{i}$ similarly to $A_\mu$ defined in terms of $a_{i}$.

The presence of $K_\mu$ leads only to minor complications of the above analysis.
All the conclusions about the number of extrema and minima remain unchanged,
however one should now distinguish symmetries of the potential and of the entire free
energy functional, see \cite{GLlong} for details.

Two local OPs also lead to the existence of quasitopological solitons.
This possibility has been known for some time;
for example, in \cite{soliton}, a soliton in the one-dimensional
two-band superconductor with a simple interband interaction term was described.
Such a soliton corresponds to the relative phase between the two condensates
that changes continuously from zero to $2\pi$
as $x$ goes from $-\infty$ to $+\infty$, and it is stable against small perturbations.
The general origin of such quasitopological solitons is obvious from the
above construction. The orbit space of all {\em non-zero} configurations of OPs
is the forward lightcone without the apex, which is homotopically equivalent to a 2-sphere $S^2$.
Depending on the exact shape of the potential, it can support either closed linear paths,
which correspond to domain walls, or closed 2-manifolds, which corresponds to strings.

{\em Multicomponent order parameters.}
In many physical situations one encounters multicomponent OPs.
Examples include 2HDM, superfluidity in $^3$He, spin-density waves, etc.
The formalism developed here works also for these cases.
Just to mention some characteristic features,
leaving a detailed discussion for \cite{GLlong}, we note
that $SU(N)$-symmetric potential depends on $N$-vectors
only via combinations $(\psi^\dagger_i \psi_j)$. Since in general
$(\psi^\dagger_1 \psi_2) (\psi^\dagger_2 \psi_1) \not = |\psi_1|^2 |\psi_2|^2$, one gets a new term
in the potential (\ref{potential}) with its own coefficient.
The definition of $r^\mu$ remains the same, but $r^\mu r_\mu \ge 0$, so $r^\mu$
can lie not only on the surface, but also in the interior of $LC^+$.
This makes definition of $B_{\mu\nu}$ unique,
fixes its eigenvalues, and depending on their signs,
one has to consider separately several cases.
Modifications to the overall results are minor,
see \cite{mink,minknew} for a 2HDM analysis.
One obtains a new phase, with $\lr{r^\mu}$ lying inside $LC^+$,
which has different symmetry properties
(in 2HDM it corresponds to a charge-breaking vacuum).
One can easily formulate conditions when it is the global minimum of the model,
so the phase diagram remains equally tractable in this case.

In conclusion, we considered the most general Ginzburg-Landau model with
two order parameters, including all possible quadratic and quartic terms
in the Landau potential. Since the minimization of the potential cannot be done
explicitly, we developed a geometric approach based on the reparametrization properties
of the model and used it to study the ground state of the model.
Future research should include dynamics of the fluctuations above the ground
state, corrections to the potential beyond the
quartic term, renormalization group flow, as well as modifications of the results
at finite temperature and in the presence of external fields.

I thank Ilya Ginzburg for useful comments.
This work was supported by FNRS and partly by grants RFBR 05-02-16211 and NSh-5362.2006.2.

\end{document}